\documentclass{iopart}

\usepackage{graphicx}
\usepackage{wasysym} % diameter symbol
\usepackage{wrapfig}
\usepackage{subfig}

\usepackage{color}

\newcommand{\geo}{GEO\,600}
\newcommand{\tem}{TEM$_{00}$}

\begin{document}
\title[Matrix Heater in GEO\,600]{Matrix Heater in the Gravitational Wave Observatory GEO\,600}
\author{H\,Wittel$^{1}$, C\,Affeldt$^{1}$, A\,Bisht$^{1}$, S\,Doravari$^{1}$, H\,Grote$^{3}$, J\,Lough$^{1}$, H\,L\"{u}ck$^{1}$, E\,Schreiber$^{1}$, K\,A\,Strain$^{2}$ and K\,Danzmann$^{1}$}
\ead{Holger.Wittel@aei.mpg.de}
\vskip 1mm
\address{$^{1}$\,Max Planck Institute for Gravitational Physics and Gottfried Wilhelm Leibniz Universit\"{a}t  Hannover, D-30167 Hannover, Germany}
\vskip 1mm
%\address{$^{2}$\,European Gravitational Observatory (EGO), I-56021 Cascina (Pi), Italy}
%\vskip 1mm
\address{$^{2}$\,SUPA, School of Physics and Astronomy, The University of Glasgow, G12\,8QQ, United Kingdom}
%\address{$^{2}$\,SUPA, School of Physics and Astronomy, The University of Glasgow, Glasgow, G12\,8QQ, UK}
\address{$^{3}$\, School of Physics and Astronomy, Cardiff University, United Kingdom}

\begin{abstract}
Large scale laser interferometric gravitational wave detectors (GWDs), such as \geo\ require high quality optics to reach their design sensitivity. The inevitable surface imperfections, inhomogeneities and light-absorption induced thermal lensing in the optics can convert laser light from the fundamental mode to unwanted higher order modes, and pose challenges to the operation and sensitivity of the GWDs.

Here we demonstrate the practical implementation of a thermal projection system which reduces those unwanted effects via targeted spatial heating of the optics.
The thermal projector consists of 108 individually addressable heating elements which are imaged onto the beam splitter of \geo. 
We describe the optimization of the spatial heating profile and obtained results.
\end{abstract}

\maketitle

\section{Introduction}
\subsection{\geo}
The gravitational wave (GW) observatory \geo\ \cite{GEO1}\cite{GEO2}\cite{GEO3} is a 600\,m long dual-recycled \cite{dual} Michelson interferometer with folded arms, located south of Hanover, Germany. 

As all other current GWDs, \geo\ uses the DC-readout method \cite{DC} for obtaining a GW measurement signal; a self-homodyne scheme, in which a photo detector measures the DC power of the output beam of \geo. The output beam however, is dominated by unwanted high order spatial modes (`HOMs') of laser light. Typically, the output beam of GEO\,600 consists of 6\,mW of \tem\ carrier light for the DC-readout, about 1\,mW of \tem\ sidebands used for controlling the interferometer, and about (depending of the alignment state of the interferometer) 30\,mW of unwanted HOMs.

\begin{figure}
	\centering
	\includegraphics[width=0.90\linewidth]{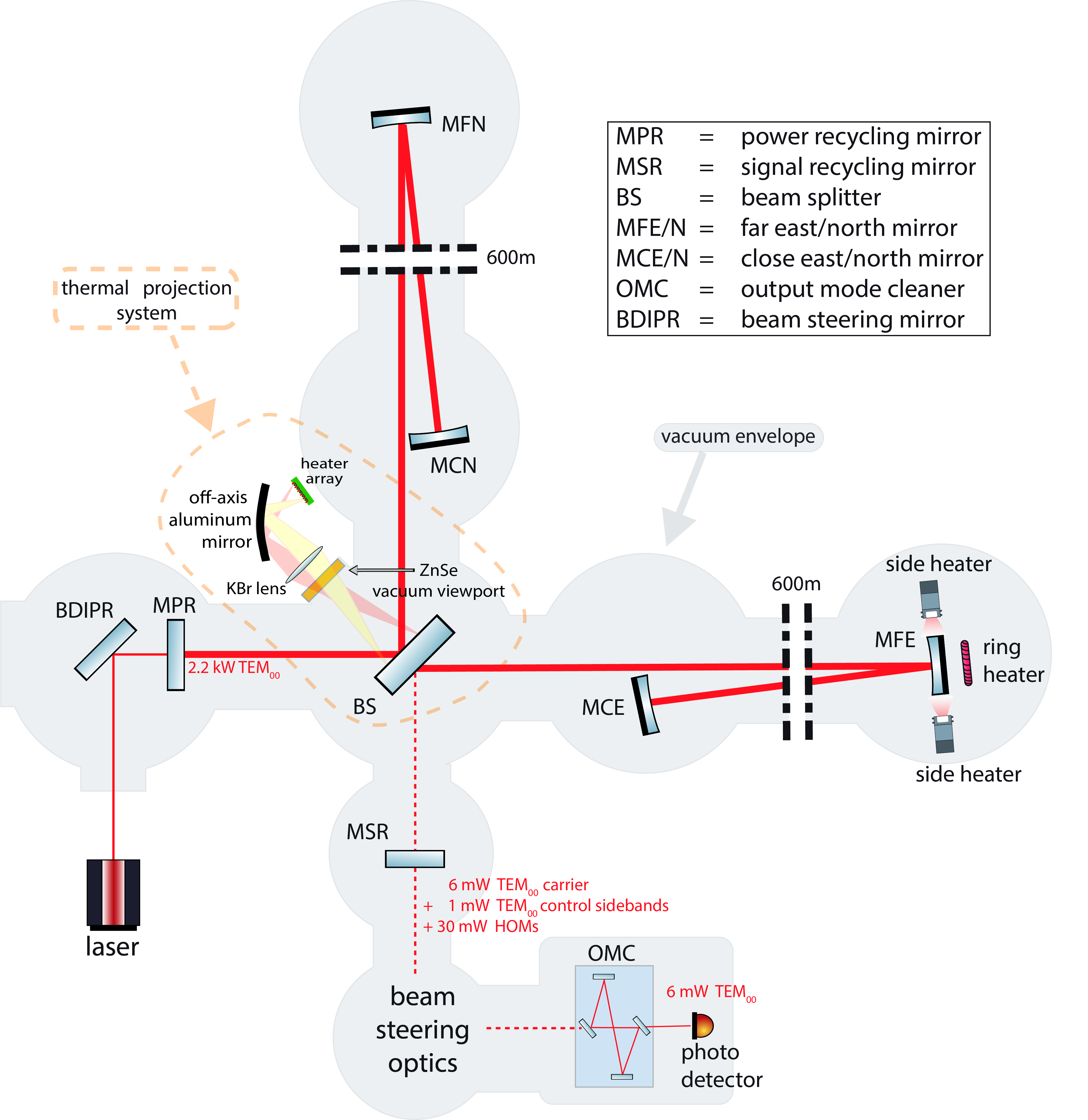}
	\caption[GEO layout]{Simplified optical layout of \geo\ including the heating setup at the far east mirror and the thermal projector at the beam splitter from this work. The vacuum enclosure in shown in light grey. Note that in this illustration the arms folded are horizontally, while in the real system they are folded in the vertical direction. All optics between the laser and BDIPR have been omitted for clarity.}
	\label{fig:geolayout}
\end{figure}

To prevent HOMs from reaching the main photo detector, a small optical cavity, the output mode cleaner (`OMC') \cite{omc} as mode selective element is placed in front of the main photo detector (see Figure \ref{fig:geolayout}). The OMC is mode-matched to the main interferometer beam and its length is controlled to be resonant to the \tem\ fundamental mode of the beam. HOMs (and control sidebands) are then reflected off the OMC. A simplified optical layout of \geo\ is shown in Figure \ref{fig:geolayout}. 

\subsection{HOMs}
Imperfect optics and thermal effects such as thermal lensing can convert \tem\ light into HOMs. Even though an OMC is an effective way preventing HOMs from reaching the main photo detector, they can still adversely affect the operation of the interferometer in several ways. For one, they can introduce spurious signals on auxiliary photo detectors which are used for the alignment of the detector subsystems and optical cavities \cite{Emil}. 

Stray light is a challenge for all GWDs. Since the spatial extent of HOMs depends on their order\footnote{it generally varies with square root of the mode order}, large order HOMs extend further than the optics of \geo. They can bounce off the (partly reflective) inside walls of the vacuum system, and a fraction of them  may eventually recombine with the main beam.

The walls of the vacuum system are not isolated from ground motion.  They are, however, partly reflective and any light that reflects or scatters from them will experience strong phase modulation. If high order modes reach the walls they will pick up varying phase shifts from the vibrating surface. A fraction of this light can eventually recombine with the main beam where the noise may affect the measurement \cite{Hildthesis}.
%In this process, phase information from the not seismically isolated vacuum system is imprinted on the beam and will affect the measurement. Even small amounts of scattered light, in the order of 10$^{-20}$\,W, can already spoil the sensitivity of \geo\ \cite{Hildthesis}.

At the limit of very large conversion of light from the \tem\ mode to HOMs, they will act as a loss channel and limit the possible power build up in the power recycling cavity, thus lowering the (usually photon shot noise limited) sensitivity.

At \geo\ we identified four main sources of HOMs.
\begin{enumerate}
	\item \textbf{Misalignment of the suspended optics.}\\
	While automatic alignment systems keep the optical system aligned over long timescales ($>$ 10\,s), at timescales of a Hertz and faster, misalignments can cause fluctuations in the power of HOMs, as can be seen by the spikes in the dark port power in Figure~\ref{fig:dpheating3}. However, the effect of misalignment is outside the scope of this work.
		
	\item \textbf{Wrong curvature of the East end mirror MFE.}\\ The radius of curvature of the mirror MFE deviates from its design value (686\,m vs designed 666\,m), and made the initial operation of \geo\ with this mirror impossible. This was corrected by installing a ring heater \cite{heater} behind this mirror. By thermal radiation it creates a thermal gradient in the mirror. Due to thermal expansion of the bulk material, the mirror's radius of curvature gets closer to the desired value.
	%would create a thermal gradient in the mirror and by thermal expansion bring it closer to the correct shape.
	
	While this made the operation of \geo\ possible, it was discovered later that the heating ring creates astigmatism in the mirror, i.e.~it curves the mirror differently in horizontal and vertical direction. 
	Additional heaters were installed \cite{side} to shape the mirror in both horizontal and vertical directions. This way it was possible to reduce the total amount of HOMs at the output port of \geo\ by 35 -- 40\%. We consider the remaining contribution to HOMs from this mirror as negligible.
	
	\item \textbf{Other imperfection of the optics, such as micro-roughness and dust particles.}\\
	Even though \geo\ uses the highest quality optics that were available at the time of installation, and is set up in a clean room environment, small imperfections and contamination on the optics are unavoidable.
	At low circulating power, when thermal effects are insignificant, we attribute all HOMs to imperfections of the optics.
	We expect these `cold' HOMs to scale linearly with the circulating power in \geo, with a scaling factor of roughly 30\,mW of HOMs at the output port per 2.2\,kW circulating power.
	
	\item \textbf{Thermal effects, in particular thermal lensing in the beam splitter.}\\
	Due to the high circulating laser power in the power recycling cavity (PRC) of \geo, thermal effects and thermally induced HOMs have to be considered. In particular the beam splitter is a strong source of thermally induced HOMs, since due to GEO's unique optical layout without arm cavities, the PRC has a very high power build up, and thus a high power passing the beam splitter substrate. Additionally, the compact layout of the central building led to an optical design placing the waist of the interferometer near the beam splitter, further increasing power density.
	As an optically transmissive element inside the PRC, the beam splitter exhibits a power dependent thermal lens which converts \tem\ light into HOMs.
	While the reflective optics in the PRC also show thermal lensing effects, we expect the beam splitter to have the largest contribution, for several reasons: In the highly reflective mirrors, absorption in the coating will introduce a bulging of the mirror surface due to thermal expansion. In the beam splitter however, we get an additional effect due to the substrate absorption and the thermal change of the refractive index\footnote{roughly ten times the size of the thermal expansion effect \cite{abs}}. Also (equal) absorption in both end mirrors would constitute a common mode effect, which would at least partly cancel in the (differential) output port.
	
	Furthermore, since the beam passes the beam splitter at an angle, the resulting thermal lens is astigmatic.
	For an increase in the circulating laser power in \geo\, we will attribute all non-linear increase in HOMs to be of thermal origin.
\end{enumerate}

\subsection{Thermal actuation}
This work investigates a method of counteracting the effects mentioned above by utilizing thermal actuation; more specifically by projecting a specific heating pattern to the beam splitter of \geo. By this means, it is possible to selectively delay areas of the laser beam wavefront, mostly due to the thermo-refractive effect. In this way -- with an appropriately shaped spatial heat distribution -- it may be possible to correct both thermal lensing effects and mirror imperfections.

Similar approaches have been investigated in the past. The heater setup at the end mirror of \geo\ has been mentioned above, but also other GWD have investigated similar approaches. The Virgo GWD used a functionally similar system (`CHRoCC') \cite{chrocc}, to correct the radius of curvature of a mirror in one degree of freedom. 

Additionally, current GWDs, such as Advanced LIGO \cite{ALIGO} and Advanced Virgo \cite{Avirgo} are facing challenges with thermal lensing in the input test masses of their arm cavities. As a mitigation, ring heaters and CO$_2$ laser projectors with masks are used to correct the laser beam wavefront \cite{ligoTCS}. However, technical noise of the CO$_2$ lasers has to be considered, hence they are used for small corrections using a mask, heating a compensation plate \emph{outside} of the highly sensitive arm cavities, while the ring heaters do the bulk of the compensation.

Different approaches that do allow arbitrary heating profiles have also been proposed:
\cite{Lawrence} discusses the use of a scanning CO$_2$ projector, and \cite{Day} provides simulations for a projector using a grid of 3$\times$3 thermal sources. Furthermore, \cite{vajente} thoroughly describes an optimization procedure of the thermal heating profile using the actuation matrix formalism. 

In this work, we present the first realization of a thermal projector for arbitrary heating profiles in a GWD. It is based on 108 thermal sources. We implemented two methods of optimizing the heating profile, and show first results.

\section{Setup}
The thermal projection system consists of an array of 9$\times$12 small heating elements located outside of the vacuum system of \geo, and an imaging system to project said array to the surface of the main beam splitter of \geo.

\subsection{Heater Array}
The heater array is a custom PCB with 108 small heating elements mounted on it. The individual heaters are re-purposed commercial platinum resistance temperature detectors (`Pt100'). Each Pt100 can produce a Planck spectrum with about 1\,W of thermal radiation (at 900\,K) in the desired wavelength range\footnote{The wavelength range that is absorbed by the beam splitter substrate material (Suprasil 311 \textregistered), i.e. longer than 4\,$\mu$m.} which is being projected onto the beam splitter.

\begin{figure}
	\centering
	\subfloat[Photograph of the heater array.]{{\includegraphics[width=0.9\linewidth]{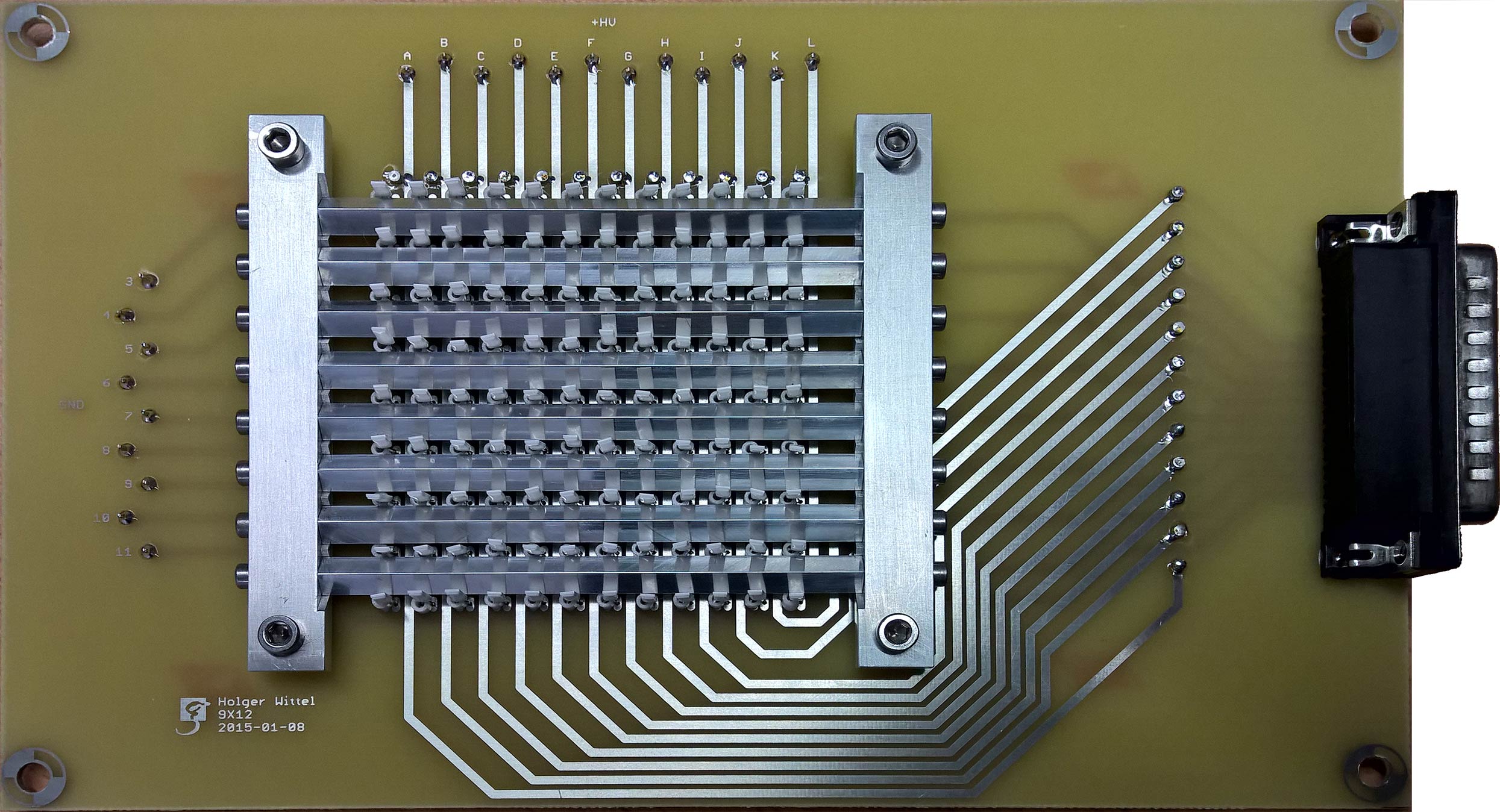}}}
	%\qquad
	\\
	\subfloat[Cross section of the heater array.]{{ \includegraphics[width=0.7\linewidth]{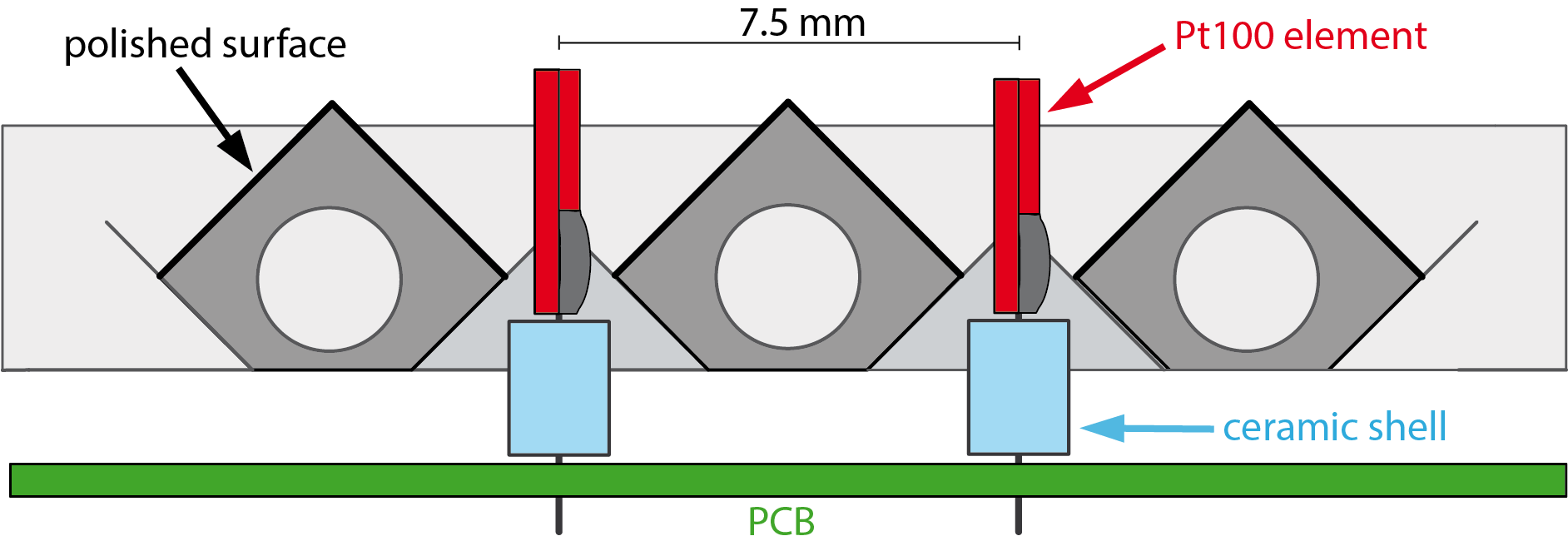}}}
	\caption{Photograph and cross section of the heater array.}
	\label{fig:9x12finalnew}
\end{figure}

The Pt100 resistors are arranged in a rectangular grid of 9$\times$12 (height$\times$width), with a center-to-center spacing (`pixel size') of 7.5\,mm $\times$ 5\,mm (h$\times$w). The heaters are standing upright on the PCB, facing angled and polished aluminum surfaces of a reflector grille, such that radiation from both flat surfaces of the Pt100s can be utilized. The setup of the heater array is shown in Figure~\ref{fig:9x12finalnew}. Each heater can be individually controlled via multiplexed driving.

The heater array is imaged to the surface of the beam splitter via the imaging system. In this case, we used an off axis parabolic aluminum mirror and a potassium-bromide (KBr) lens, which project the image through a zinc-selenide (ZnSe) vacuum-window onto the beam splitter. All of the materials for the optics were chosen for good (broadband) transmission of thermal radiation from the heater array.
The projection system is chosen to have a magnification factor of two.  The throughput of the optical system is limited by the solid angle between the vacuum viewport and the beam splitter to a numerical aperture NA of 0.06.

\section{Procedure}
As a first step in the operation of the thermal projection system, it is necessary to align the projected image of the heater array to the beam path in the beam splitter of \geo. 

The scheme we have devised for doing this works by designating a specific heater as `center pixel' and modulating its driving current, which will result in modulated (spatially limited) heating on the beam splitter. Due to the $dn/dT$ effect, the same modulation will couple into the differential arm length signal of \geo, depending on the overlap of the heated area with path of the main laser beam. In the alignment process, the overlap of the central pixel with the laser is maximized. 

\begin{figure}
	\centering
	\includegraphics[width=0.65\linewidth]{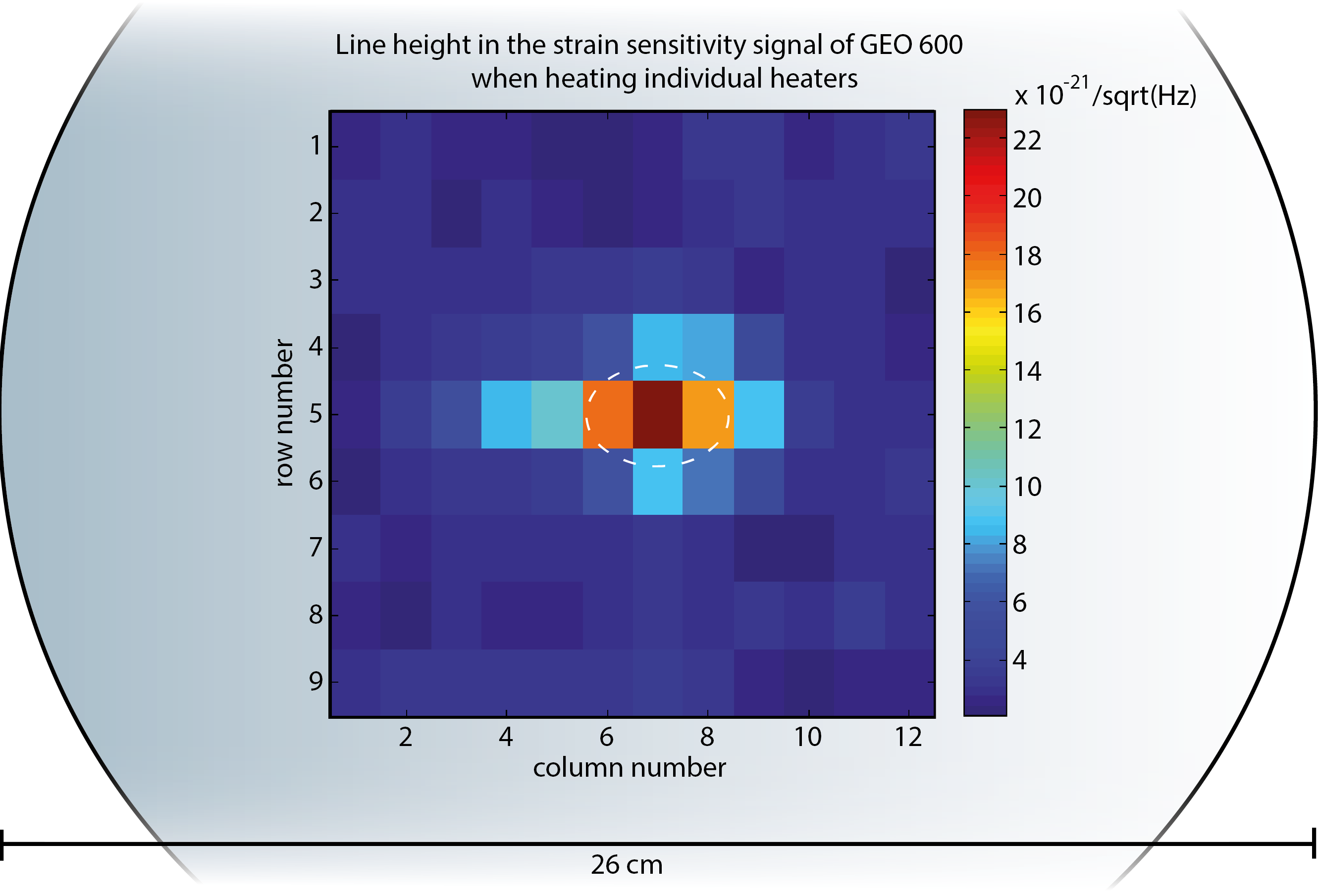}
	\caption{Measured overlap of the projected pixels of the heater array and the fundamental TEM$_{00}$ mode of the main laser beam in \geo. For comparison, the (1/e$^2$ in intensity) size of the laser beam spot on the beam splitter is represented via white dashes, while the outline of the beam splitter is depicted in the background.}
	
	%\caption{}
	\label{fig:alignnew}
\end{figure}

Furthermore, we can use this method to map out the overlap of each individual pixel with the fundamental \tem\ mode of the laser beam in the beam splitter, as shown in Figure~\ref{fig:alignnew}. As expected, due to the laser beam hitting the beam splitter at an angle close to 45 degrees, the beam profile appears to be oval.

\begin{figure}
	\centering
	\includegraphics[width=1\linewidth]{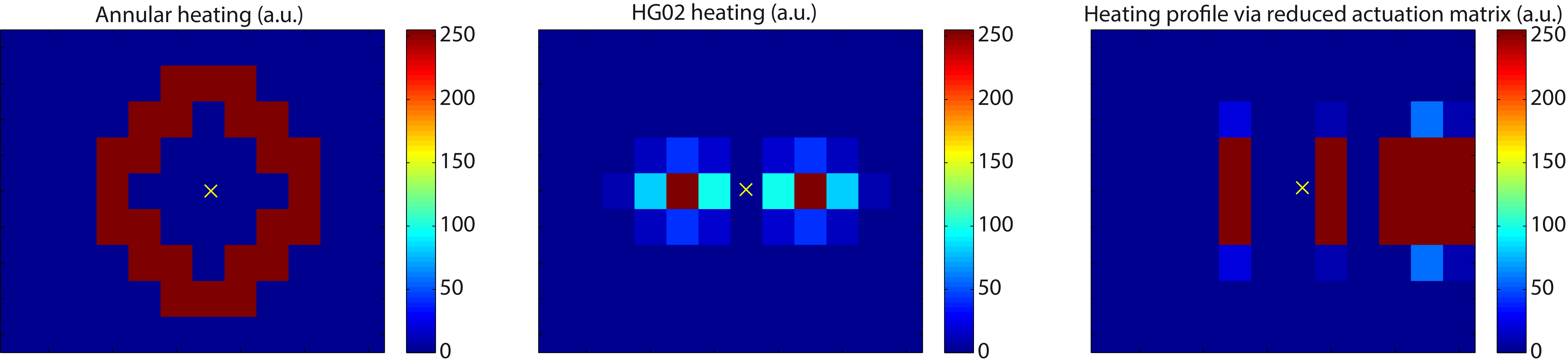}
	\caption[heating patterns]{The heating patterns used in this work (intensity in a.u.). The yellow `x' marks the center of the laser spot on the beam splitter.\\
		\begin{tabular}[]{ll} 
			left:   & annular heating profile\\
			center: & HG$_{20}$ heating, to counteract HG02 mode\\ 
			right:  & heating profile against HG modes 2,7 and 8, via reduced actuation matrix\\
		\end{tabular}
		%left: HG$_{20}$ heating,\\center: best naive annular heating profile,\\right: heating profile obtained via reduced actuation matrix.\\
}
	\label{fig:patterns}
\end{figure}

Finally we try to reduce the amount of HOMs produced in \geo. For this, it is necessary to find a suitable setting for each heating element (`heating pattern'). Using an approach in which individual heaters are adjusted one by one can be tedious, due to the large number of degrees of freedom and the long thermal timescales.  
\\
We followed two approaches to determine a suitable heating pattern:

\textit{1. The naive approach:}
Since a large fraction of the HOMs in \geo\ originates from thermal effects/ thermal lensing at the beam splitter, it should be possible to mitigate this issue by an annular heating pattern, which will create a negative thermal lens. Overall the aim is to flatten the thermal gradient caused by the high powered laser beam in the beam splitter.
%Via trial and error we determine the best (using the total power of HOMs at the output port of \geo) annular heating pattern. 
Via trial and error, using the total power of HOMs at the output port of \geo\ as a measure, we a suited annular heating pattern. The result is shown in Figure~\ref{fig:patterns}.

\textit{2. The actuation matrix approach:}
As a second way to find a suitable heating pattern, we employed a more deterministic technique, as described in \cite{vajente}. This technique works by defining a base-set of heating patterns. Each pattern of the base set is applied, and the effect on HOMs is recorded. The relation that is obtained this way can be expressed in matrix $A$ in the form:
\[ 
\vec{HOM} = \textbf{A} \, \vec{h},
\]
with the $\vec{HOM}$ being the power in the HOMs, and $\vec{h}$ being a vector of the basis heating patterns.

Once the matrix $\textbf{A}$ is known, it can be inverted, and one can obtain a linear combination of the basis heating patterns which produces a desired distribution of HOMs\footnote{Here we are only interested in the special case of minimizing the amount of HOMs}. A reasonable basis set of heating profiles may be the phase profile of the high order Hermite-Gauss (HG) modes that we want to affect. We expect especially the second order HG modes to be of importance, since a mode mismatch due to a thermal lens in the beam splitter would mostly produce these modes.

Due to several practical challenges, we had to adapt the procedure in \cite{vajente}, which we will describe in the following. 

In theory one would use a bias in the heating, i.e. choosing half power for all heaters as zero point, as this would allow for `negative' heating. In our setup however, we noticed an increase in HOMs with all heaters at a constant power, which we attribute to a not perfect fill factor of the heating. 
Consequently, we had to slightly change the basis set of heating profiles; We still use heating profiles based on the phase profile of the high order Hermite-Gauss (HG). For each HG mode, we only use areas of positive phase as one base vector, and use the areas of negative phase as a second base vector.

Furthermore, in order to measure the impact of each heating profile in our chosen base to the HOMs, we use the OMC at the output port of \geo. In normal operation its length is adjusted to be resonant to the \tem\ mode. By changing its length, we can make it resonant for HOMs of different orders instead, and measure their power on the main photo detector. A drawback of using the OMC as measurement tool for HOMs is that it is mode-degenerate; all modes of the same order will resonate at the same length, i.e. an HG$_{01}$ mode cannot be distinguished from an HG$_{10}$ mode with this method. Therefore, this method will work best for mode orders which are dominated by a single HOM. In \geo\ this is the case for mode orders 2, 7, and 8.

\section{Results}
\begin{figure}
	\centering
	%\includegraphics[width=1\linewidth]{figures/dp_heating3.pdf}
	%\caption[Dark port power]{Plot of the time series of the power at the dark port at GEO\,600 (dominated by HOMs, with $<$\,2\,mW \tem\ content) at normal and increased (1.6\,h~--~4.5\,h) operating power. Periods during which the thermal projector is used to suppress HOMs are numbered and marked by the colored background.}
	\subfloat[Dark port power.]{{\includegraphics[width=0.97\linewidth]{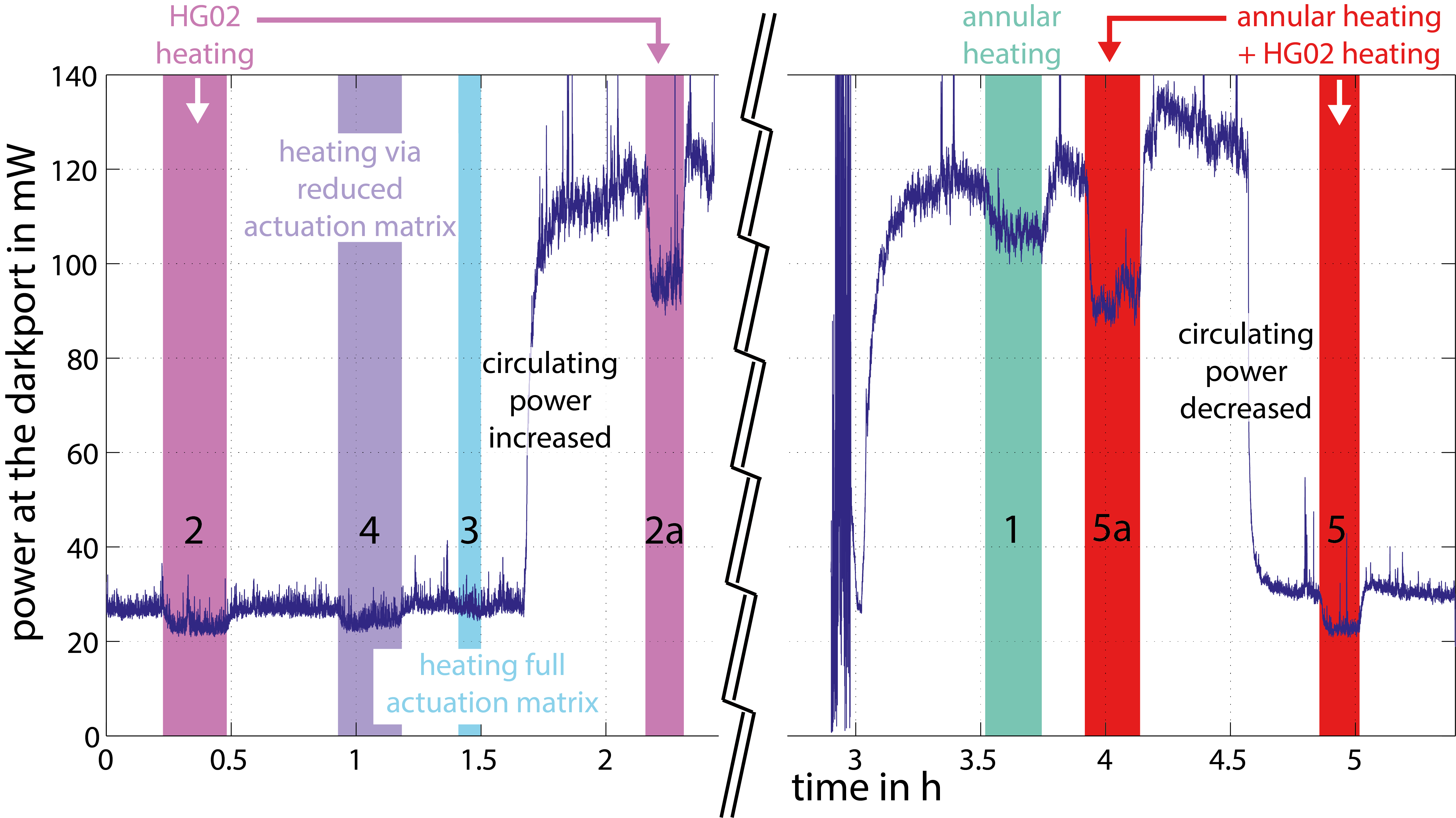}}}
	\\
	\subfloat[Cutout view of select areas from above.]{{ \includegraphics[width=.955\linewidth]{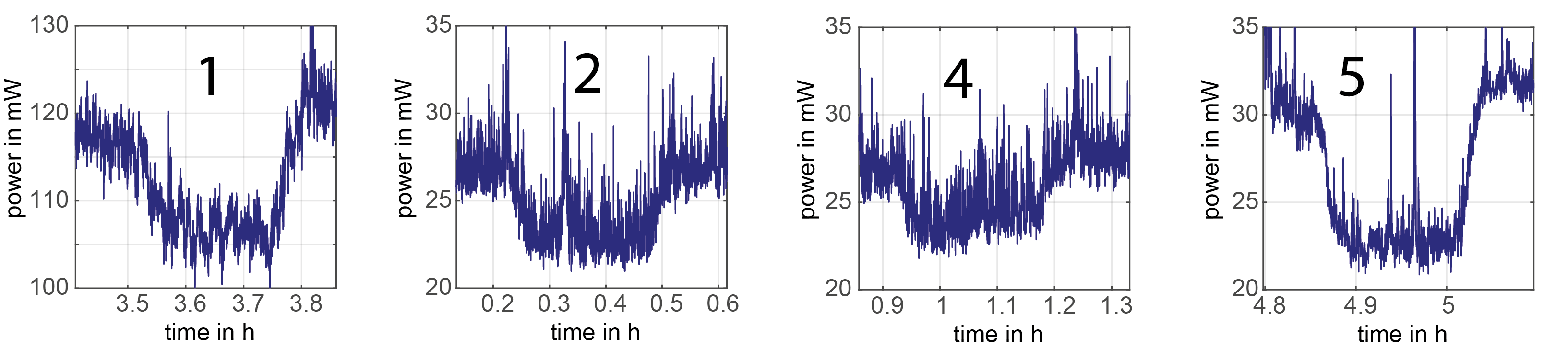} }}
	
	\caption[Dark port power]{Plot of the time series of the power at the dark port at GEO\,600 (dominated by HOMs, with $<$\,2\,mW \tem\ content) at normal and increased (1.6\,h~--~4.5\,h) operating power. Periods during which the thermal projector is used to suppress HOMs are numbered and marked by the colored background.}
	\label{fig:dpheating3}
\end{figure}

%6247
%6214
We tested the two approaches from above, and judge the outcome by the effect on total power at the output port of \geo, which is dominated by HOMs. Figure~\ref{fig:dpheating3} shows a time series of the power at the output port of \geo\ at the standard operating power (2.2\,kW circulating power in the PRC) and with an increased circulating power (3.5\,kW) to increase the influence of the thermal lensing effect. 
Note that for the experiments depicted in Figure \ref{fig:dpheating3}, the \tem\ content in the output beam was intentionally reduced to be $<$\,2\,mW, to make the HOMs (even more) dominate the power at the output port of \geo. Furthermore, the \tem\ power is kept constant, any changes in the output power can be attributed to HOMs.

With the annular shape, based on the naive approach, we obtain an improvement in HOM power in the order of 10\% (`1' in Figure \ref{fig:dpheating3}). 

As a next step we apply a single heating profile from the chosen basis set, with a shape similar to the HG$_{02}$ modes (`2' and `2a' in Figure \ref{fig:dpheating3}), since we expect this mode to be affected the most when raising the interferometer power and it is one of the strongest modes even at lower power. With this we achieve an improvement in the order of 15\% (standard power) and 20\% (increased power).

We also tested the heating profile obtained by the full actuation matrix (`3' in Figure \ref{fig:dpheating3}), but do not achieve a significant improvement in HOMs. We attribute this to the fact that many of those mode orders are not dominated by a single HOM and therefore the actuation matrix based on the (mode degenerate) OMC may not be accurate for the degenerate mode orders. When reducing the actuation matrix to the mode orders dominated by a single mode (2,7,8 in GEO, `4' in the Figure), we obtain results similar to the HG$_{02}$ heating profile. Furthermore, by solving the inverse actuation matrix, we determine that the required heating profile contains elements that are greater than the maximum power that the heater array can apply. Therefore, we combined this heating profile with the best annular one to increase the total power transferred to the beam splitter (by involving more heater elements). This results in an improvement in HOMs of 31\% in the high- and 24\% in the standard power state (`5a' and `5' respectively in Figure \ref{fig:dpheating3}).

% this plot was made in Het with reduces SBs! It shows mostly only HOMs.

\section{Summary, Outlook and Discussion}
We have demonstrated the use of a thermal projector for the generation of arbitrary heating profiles in a large scale GWD, and achieved a reduction of unwanted HOMs, and therefore an improvement in interferometer contrast by 30\%. 
One finding is that more power delivered to the beam splitter by the thermal projection system may be beneficial. Therefore, the thermal projection system is currently being upgraded; The optical system outside of the vacuum chamber has been replaced by an in-vacuum lens, which due to the increased opening angle, will allow for a factor of five increase in the power transmission. Furthermore, the heater matrix in this work used a row-by-row multiplexing for driving the heaters, which is easier to build, but produces a signal in \geo\ at the multiplexing frequency. A new layout will work without multiplexing and instead provide an individual channel for each heater. 

While the method of determining an ideal heating profile has been shown to work, the measurement of the exact actuation matrix poses a challenge with the existing infrastructure (i.e. with the mode degenerate OMC). An alternate approach could be to use the obtained heating profile for the well-defined HOMs, and use it as a starting point for an in-situ optimization, for example via the Newton method. Global optimization methods searching beyond local maxima, as e.g. simulated annealing or genetic algorithms, may be useful as well.
%\clearpage

\ack{We thank the GEO collaboration for the development and construction of 
GEO\,600. The authors are also grateful for support from
the Science and Technology Facilities Council (STFC),
the University of Glasgow in the UK,
the Max Planck Society,
the Bundesministerium f\"ur Bildung und Forschung (BMBF),
the Volkswagen Stiftung,
the cluster of excellence QUEST (Centre for Quantum Engineering and Space-Time Research),
the international Max Planck Research School (IMPRS),
and the State of Niedersachsen in Germany.

%This work has been performed with the support of the European Commission under the Framework
%Programme 7 (FP7) `Capacities', project \emph{Einstein Telescope} (ET) design study 
%(Grant Agreement 211743) (\tt http://www.et-gw.eu/)
This article has been assigned LIGO document number LIGO-P1800116.
}

\end{document}